\documentclass[prl,twocolumn,aps]{revtex4}
\usepackage[T1]{fontenc}
\usepackage[latin1]{inputenc} 
\usepackage{epsfig}
\usepackage{amsmath}
\usepackage{amssymb}
\usepackage{amsthm} 
\usepackage{bm}
\newcommand{\ket}[1]{\left|#1\right>}

\newcommand{\nn}{\nonumber\\}

\newcommand{\bea}{\begin{eqnarray}}
\newcommand{\ea}{\end{eqnarray}}
\newcommand{\eea}{\end{eqnarray}}
\newcommand{\ord}{{\cal O}}
\newcommand{\sumint}[1]

\begin{document}

\title{Fragmented many-body ground states for scalar bosons
in a single trap}

\author{Philipp Bader and Uwe R. Fischer}

\affiliation{Eberhard-Karls-Universit\"at T\"ubingen,
Institut f\"ur Theoretische Physik\\
Auf der Morgenstelle 14, D-72076 T\"ubingen, Germany}

\begin{abstract}
We investigate whether the many-body ground states of bosons   
in a generalized two-mode model with localized inhomogeneous 
single-particle orbitals and anisotropic long-range interactions (e.g. dipole-dipole interactions), are coherent 
or fragmented. It is demonstrated that 
fragmentation can take place in a single trap for positive values of the 
interaction couplings, implying that the system is potentially stable.  
Furthermore, the degree of fragmentation is shown to be insensitive
to small perturbations on the single-particle level. 

\end{abstract}


\maketitle

Fragmentation in many-body states of interacting bosons is defined 
by the single-particle density 
matrix having more than one macroscopic eigenvalue 
\cite{Penrose,Pethick}, 
leading to macroscopic occupation of more than one state. 
In the continuum, and for 
repulsive contact interactions,
the Fock term in the total energy favors a single 
condensate;  
a negative contact interaction   
implies the collapse of the gas before ground-state 
fragmentation sets in \cite{Nozieres}. 
On the other hand, in spatially well-separated systems, 
fragmentation can be obtained for deep double wells \cite{Spekkens}, 
or in its periodic extension Mott state in optical lattices \cite{Greiner}. Further examples for fragmentation 
have been found \cite{Mueller}, 
e.g., when internal degrees of freedom and sufficient
symmetry in the interaction are present \cite{HoYip},
in the Richardson pairing model \cite{Dukelsky},
in the crossover to fermionization and for excited states 
\cite{Cederbaum08}, and in rapidly rotating gases \cite{Wilkin,remark}. 
However, no 
ground-state fragmentation of a scalar 
Bose gas in a single trap has been reported.  

So far, only contact interactions between the bosons
and quasi-homogeneous or spatially periodic systems  
were considered. The question then arises whether for general 
inhomogenity, nonlocality and anisotropy, 
i.e. given the localization in 
a single trap with inhomogeneously distributed single-particle orbitals (modes) 
and for partly positive and negative long-range interactions (of which the archetype 
is the dipole-dipole interaction), ground-state fragmentation can occur. 
In the following, we consider for simplicity and easy comparison 
with the double-well case,  
that only two single-particle orbitals are occupied 
\cite{Milburn}. 
Examples are gases strongly confined 
in a given direction, 
e.g., close to the quasi-2D limit; 
the question then arises whether fragmentation occurs 
with respect to that strongly confining direction. 
A two-mode approximation has, furthermore, 
the important benefit that the many-body states of the system
can be found essentially exactly numerically (and for special cases
analytically). We demonstrate that by tuning the four 
interaction matrix elements relative to each other, 
various many-body states 
-- coherent and fragmented states, as well as coherent superpositions of degenerate 
macroscopically distinct quantum states \cite{Schroedinger} -- can be accessed in a single trap.
In addition, the fragmented states are 
not susceptible to decay to a non-fragmented (coherent) state 
because of a perturbation on single-particle level \cite{Kanamoto,Jackson}, 
due to the fragmentation being based to the 
values 
of the interaction couplings.

We begin with a general quadratic plus 
quartic Hamiltonian for two interacting modes, 
\bea 
\hat H&=&\epsilon_0\hat a_0^\dagger\hat a_0
+\epsilon_1\hat a_1^\dagger\hat a_1 
- \frac\Omega2 \left( \hat a_0^\dagger\hat a_1 + {\rm h.c.} \right) \nn 
& & +\frac{A_1}2 \hat a^\dagger_0\hat a^\dagger_0 \hat a_0 \hat a_0 
+\frac{A_2}2 \hat a^\dagger_1\hat a^\dagger_1\hat a_1\hat a_1\label{Ha}\\
& & +\frac{A_3}{2}\left(\hat a_0^\dagger\hat a^\dagger_0
\hat a_1\hat a_1+ {\rm h.c.}\right)  
+\frac{A_4}2 \hat a_1^\dagger \hat a_1 
\hat a_0^\dagger\hat a_0 .\nonumber
\ea 
The interaction coefficients are given by   
$A_1= V_{0000}, A_2 = V_{1111}, 
A_3 = V_{1100}=V_{0011}$ (taken to be real), 
and  $A_4 = V_{0101}+V_{1010}+V_{1001}+V_{0110}$, where 
$V_{ijkl} = \int d^3r
\int d^3r'\Psi^*_i({\bm r})\Psi^*_j({\bm r}')
V_{\rm int} ({\bm r}-{\bm r}') \Psi_k({\bm r}')\Psi_l({\bm r}) $. 
It is of major importance for the discussion to follow 
that we include pair-exchange between the two modes 
due to scattering of pairs of bosons $\propto A_3$,  
in addition to the standard density-density
type terms $\propto A_1,A_2,A_4$. We stress that 
this term is {\em absent} in the homogeneous continuum due to momentum conservation.  
We also note 
that the independence of the coefficients $ A_i $ makes 
\eqref{Ha} different from rotationally invariant  
interaction Hamiltonians in gases with internal degrees of freedom, 
where the modes are components of a spinor 
\cite{Mueller,HoYip}. 
For Fourier transformable interaction potentials we have, by convolution, 
$V_{ijkl} = 
\frac1{(2\pi)^3} 
\int d^3 k \tilde\rho_{il}(-{\bm k}) \tilde
V_{\rm int} ({\bm k})\tilde\rho_{jk}({\bm k}) $, where the 
$\tilde\rho_{jk}({\bm k})$ are Fourier
transforms of $\Psi^*_j({\bm r})\Psi_k({\bm r})$. 
The single-particle energies 
$\epsilon_i = \int d^3r [ \frac{\hbar^2}{2m} |\nabla\Psi_i|^2 + V_{\rm trap} |\Psi_i|^2 ]$ 
are given by kinetic plus trap contributions.  
Finally, the Josephson-type coupling 
$\Omega$ connects the single-particle states 
and corresponds to a tunneling rate \cite{Leggett,Omeganote}. 

We emphasize that differing magnitudes and signs of the $A_i$
can stem both from the interaction  
{\em and} the form of the single particle orbitals.  
This only simplifies in the 
continuum limit (the orbitals are plane waves then), 
where 
the interaction couplings directly reflect anisotropy and long-range character 
of the interaction.  Note that for contact interactions, 
{all} $A_i$ negative necessarily implies collapse 
for large systems, 
occurring by the fact that choosing ever more singular
single particle orbitals is energetically advantageous.
On the other hand, by 
adjusting the single-particle orbitals
(the trapping) in conjunction with in general anisotropic interactions like the   
dipole-dipole interaction, the magnitude and sign of the interaction 
coefficients $A_i$ can be engineered \cite{dipole-dipole}.
The Hamiltonian \eqref{Ha} therefore represents a minimal model
to investigate whether ground-state fragmentation 
takes place for potentials and traps with arbitrary anisotropy and shape, 
leading to essentially independent interaction couplings $A_i$. 
In a single trap, the two terms in the last line of \eqref{Ha} are
important, in distinction to the double-well
case ($\hat a_0,\hat a_1$ 
are then the annihilation operators for particles in the left and right wells, 
respectively), where they both are exponentially suppressed. 
In particular, we will see that the pair-exchange term $\propto A_3$ 
decides upon the class of many-body solutions obtained.

Let $|l\rangle  \equiv 
|N-l,l\rangle $ be a $N$-particle two-mode state, with 
$N-l$ in the ground and $l$ particles in the excited state. 
Then, all many-particle states can be expressed by 
$|\Psi\rangle=\sum_{l=0}^N\psi_l|l\rangle$. 
The energy reads with the Hamiltonian 
\eqref{Ha}, 
\begin{multline}
\langle \Psi|\hat H|\Psi \rangle
=\sum_{l=0}^N
\left[
\frac{A_3}2 d_l (\psi_l^*\psi_{l+2} + \psi_l \psi^*_{l+2}) \right. \\
\left.-\frac{\Omega}2 \omega_l (\psi_l^*\psi_{l+1} + \psi_l \psi^*_{l+1})
+ c_l |\psi_l|^2  \right],  \label{functional}
\end{multline} 
where the diagonal coefficient 
$c_l = \epsilon_0(N-l)+\epsilon_1 l+\frac12A_1(N-l)(N-l-1)
+\frac12A_2 l(l-1)+\frac12A_4 (N-l)l$, 
$d_l= \sqrt{(l+2)(l+1)(N-l-1)(N-l)}$, while
$\omega_l = \sqrt{(N-l)(l+1)}$. 
Differentiating the functional \eqref{functional} with respect to 
$\psi_l$, the linear system to be solved reads 
\begin{multline} 
\langle l|\hat H|\Psi\rangle = E\psi_l = 
\frac{A_3}{2}(d_l\psi_{l+2}+d_{l-2}\psi_{l-2}) \\
\hspace*{4em}-\frac\Omega2 (\omega_l \psi_{l+1} +\omega_{l-1} \psi_{l-1}) +
c_l\psi_l \,.\label{diffeq}
\end{multline} 
There are three important cases,  
$A_3<0$, $A_3>0$, $A_3=0$, for a given set of interaction couplings 
$A_1,A_2,A_4$ (which enter $c_l$), assuming $\Omega\ge 0$.

To get first insight, and introduce some notions, 
we begin by discussing the simplest case of 
no pair-exchange between orbitals and no coupling of levels,  
$A_3 = \Omega=0$. 
For this configuration, exactly one state $|l\rangle$
will obtain, i.e., a definite occupation of both the 
ground and excited single-particle state, 
because the Hamiltonian is diagonal in $l$. 
The necessary condition for such a 
{\em fragmented} state is convexity of the energy parabola in $l$, 
$\frac{\partial^2 c_l}{\partial l^2}=A_1+A_2-A_4>0$. The energy is 
minimized for 
$l_{\rm min}=\frac N2 + \frac{(A_1-A_2) (N-1) 
+ 2(\epsilon_0 - \epsilon_1)}{2(A_1 + A_2 - A_4) }
$. 
On the other hand, if the energy parabola is concave, we get a condensate, i.e.
all particles accumulating in one of the 
two-mode states $|N,0\rangle$ $(l=0)$, $|0,N\rangle$ ($l=N$), 
depending on which of the two states 
has lower energy. There is however one 
exceptional case. If the energies of the two states 
are identical, leading, by $c_0=c_N$,
 to the condition 
 $\epsilon_1-\epsilon_0 = (N-1)(A_2-A_1)/2$,  
concavity leads to the coherent superposition of the macroscopically distinct
and degenerate many-body states $|N,0\rangle$ and $|0,N\rangle$ 
\cite{Mueller,Ciobanu,Cirac}, 
characterized by 
number fluctuations of, say, the lower single-particle 
state, $\Delta N_0^2 =\langle \hat N_0^2\rangle-\langle \hat N_0 \rangle^2 = 
\ord(N^2)$.  

\begin{center}
\begin{figure}
\centering
\hspace*{-1.2em}
\vspace*{-1em}
\includegraphics[width=.475\textwidth]{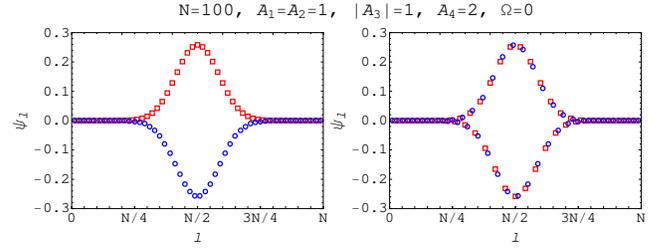}
\caption{\label{Omega=0} Numerical solutions of \eqref{twoeqns} for the 
indicated parameter set, for $A_3=-1$ (left) and $A_3=1$ (right), 
centered around $l=N/2$ (the $A_i$ are given in an arbitrary overall 
unit of energy). 
The red squares indicate the even $l$ 
sector (distribution of the $\phi_l$), 
and the blue circles the odd $l$ sector (distribution of the $\Phi_l$), 
using the ground-state superposition $|\phi \rangle - |\Phi \rangle $ for 
clarity of representation.}
\end{figure}
\end{center}

In the case of a still vanishing $\Omega=0$, but with a general set of 
$\{A_i\}$, Eq.\,\eqref{diffeq} 
decomposes into two independent equations 
connecting 
the even and odd $l$ sectors of $\psi_l$, 
\begin{multline}
E\psi_{2k}=c_{2k}\psi_{2k}+d_{2k}\psi_{2k+2}+d_{2k-2}\psi_{2k-2},\\ 
\label{twoeqns}				
E\psi_{2k+1}=c_{2k+1}\psi_{2k+1}+d_{2k+1}\psi_{2k+3}+d_{2k-1}\psi_{2k-1}, 
\end{multline}
$\forall\,k\in \{0,\dots,N/2\}$. 
We have numerically established that (to faster than exponential 
accuracy in $N$, i.e. $\propto\exp[-N^\alpha]$ with $\alpha >1$) 
the two resulting ground states in the sectors of even and odd $l$, 
$\ket{\phi}=\sum_{l}\phi_{l}\ket{2l}$ and 
 $\ket{\Phi}=\sum_l\Phi_l\ket{2l+1}$ 
are degenerate when the continuum limit (see below) is valid. 
We display typical results for $\psi_l$ 
distributions in Fig.\,\ref{Omega=0}, where we put 
$\epsilon_0=\epsilon_1=0$ here and in the plots to follow 
(the $\epsilon_i$ influencing the location of the maximum 
of the $\psi_l$ distribution, see Eq.\,\eqref{shift} and 
the discussion following it).
\vspace*{-2.5em} 

When the pair-exchange coupling $A_3$ is positive (Fig.\,\ref{Omega=0}, right panel), 
it causes alternating signs, $\mathrm{sign}(\psi_l\psi_{l+2})=-1$, induced by
its occurrence in the energy functional (2): The energy is minimized by
choosing $A_3\psi_l\psi_{l+2}<0$.
Using the same argument, $A_3$ negative (Fig.\,\ref{Omega=0}, left panel)
requires the signs of the $\psi_l$ to be identical. 
The energy contribution of the pair-exchange term is generally 
negative, and its maximization forces $\psi_l$ to be real.

The existence 
of sign changes has profound effects on the properties of
superposed ground states.
The first order coherence 
$g_1 = \frac12 \langle \hat a_0^\dagger \hat a_1 
+ \hat a_1^\dagger \hat a_0\rangle$ vanishes for any superposition
$a\ket{\phi}+b\ket{\Phi}$ in the presence of oscillations in the even and odd
sectors of $\psi_l$, i.e. for $A_3>0$, 
leading to well-defined fragmented ground states, in 
the sense that their degree of fragmentation $\mathfrak F$
in \eqref{Fragmentation}
is independent of the concrete realisation via the weights $a$ and $b$. 
To the best of our knowledge, this mechanism of fragmentation for positive
pair-exchange coupling in a trapped scalar Bose gas 
has not been discussed before in the literature. 
For $A_3$ negative, on the other hand,  
the possibility of superpositions allowed by the ground state
degeneracy causes the system to be arbitrarily tunable between fully
fragmented and coherent states; this can only be avoided by lifting
the degeneracy by a small but finite $\Omega$.

We define the degree of fragmentation to be 
$\mathfrak{F} = 1-{|\lambda_0-\lambda_1|}/N$,
where $\lambda_{0,1}$
are the eigenvalues of the single-particle density matrix 
$\rho^{(1)}_{\mu\nu}=\langle a^\dagger_\mu a_\nu\rangle$, given by
$ \lambda_{0,1}/N = \frac12 \pm \sqrt{\frac14 -\frac1{N^2} (N_0 N_1  
- |\langle \hat a_0^\dagger \hat a_1 \rangle|^2) } $. 
They collapse into a single nonvanishing one,
$\lambda =N$, as soon as the state becomes coherent, 
$\langle \hat a_0^\dagger \hat a_1 \rangle = \sqrt{N_0N_1}$, and therefore 
$\mathfrak F\equiv 0$ for coherent states.
The degree of fragmentation evaluates to 
\bea
\mathfrak{F} = 1-\frac{2}N \sqrt{\left|\langle \hat a_0^\dagger \hat a_1 \rangle\right|^2 +\left(\frac N2 
-\sum_{l=1}^N|\psi_l|^2 l \right)^2}\,,   \label{Fragmentation}
\ea
where $\langle \hat a_0^\dagger \hat a_1 \rangle= \sum_{l=1}^N \psi_{l-1}^*\psi_l \sqrt{l(N-l+1)}$. 

\begin{center}
\begin{figure}
\centering
\includegraphics[width=.4\textwidth]{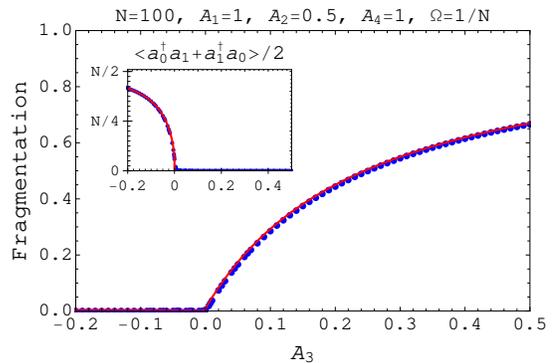}
\caption{\label{Fragment} The degree of 
fragmentation $\mathfrak{F}$ in Eq.\,\eqref{Fragmentation} 
across the transition from coherent to fragmented states
at $A_3=0$, for small but finite $\Omega = 1/N$.
The blue dots represent the fragmentation from 
the numerical solution of \eqref{diffeq}; 
the solid red line is the continuum approximation 
from \eqref{FragAnal} in the $\Omega\rightarrow 0$ limit
[the agreement, though not visible
on the scale of the figure, has also been verified for the $A_3<0$ side].
The inset shows the first-order coherence, with the 
continuum result 
$g_1=  
 \sqrt{N^2/4 -\mathfrak{S}^2} \exp[-1/(4N\sqrt R)]$.} 
\end{figure}
\end{center}

The degeneracy of the two many-body states is lifted by $\Omega$, which 
connects $\psi_l$ and $\psi_{l\pm1}$, and 
resolves this degeneracy already for $\Omega= \ord (1/N)$ 
[so that the energy contribution of the $\Omega$ term is still
suppressed to $\ord(1/N^2)$ compared to the $A_3$ term]. 
For $A_3<0$, turning on $\Omega$ then establishes a coherent state
characterized by number fluctuations $\Delta N_0^2 \propto N$ (with a 
proportionality factor of order unity), while 
for $A_3>0$, fragmentation persists for $\Omega > 0$.
We illustrate the change of $\mathfrak F$ with $A_3$, for finite $\Omega$,  
in Fig.\,\ref{Fragment}. We have checked the robustness of the fragmentation 
obtained on the $A_3>0$ side upon increasing the interlevel
coupling up to $\Omega\sim \ord(N^0)$, as well as
for variations of the values of the single-particle energies 
$\epsilon_0$ and $\epsilon_1$ to the same order.
The degree of fragmentation 
is therefore stable for small perturbations 
on the single-particle level, different to what was found in 
\cite{Kanamoto,Jackson}, where the origin of fragmentation 
is distinct from our interaction-couplings based mechanism.
Finally, because the energy contribution of the $A_3$ term is negative,   
for a concrete realization with sufficiently small $A_3$  
the fragmented state has to be at a (local) 
minimum of the energy in the parameter space of the orbitals $\Psi_i$. 
The latter may be determined by a variational ansatz, 
e.g., by using the ellipsoid half-axes of harmonic oscillator
trial wave functions as the orbital parameters \cite{dipole-dipole,Elgaroy}. 
 
\vspace*{-2.5em} 

We now consider the continuum limit of slowly varying $\psi_l$ 
\cite{Spekkens}, 
which allows for an analytical solution in the $\Omega\rightarrow 0$ limit.
Employing the definition $j\equiv l - \frac N2 $,
and expanding $c_j,d_j$ to quadratic order in $j/N$, 
neglecting $\ord (1/N)$ terms, we obtain a
differential equation  for 
$|\psi(j)|$, 
\begin{multline}
\left[-\frac{|A_3| {N^2}}2  \frac{\partial^2}{\partial j^2} 
+ 
\frac{|A_3|}{2R(\{A_i\})} 
\left(j-\mathfrak{S}\right)^2 \right]
|\psi| 
= E' |\psi|\,. \label{Schroedinger} 
\end{multline} 
Here, $E'$ differs from $E$ by a 
constant, and  
the location of the distribution center, the {\em shift}, 
is given by  \cite{note}
\bea
\mathfrak{S} =\frac{N(A_1-A_2)/2 
+\epsilon_0-\epsilon_1}{A_1+A_2+2|A_3|-A_4} .
\label{shift}
\ea
The solution of lowest energy of the 
equation for $|\psi(j)|$ in \eqref{Schroedinger} 
is a Gaussian centered at $\mathfrak{S}$, cf.\,\,Fig.\,\ref{Omega=0}, 
and of width
$
\sigma_{\rm osc}  
= \sqrt{N} 
R^{1/4}(\{A_i\})
$, 
provided that  
\bea
R(\{A_i\})= \frac{|A_3|}{A_1+A_2+2|A_3|-A_4} \label{Rdef}
\ea 
is positive. 
Note that particle number [to $\ord (1/N)$] and
single-particle energies do not enter $R$,  
in distinction to the corresponding criterion in 
the double-well case \cite{Spekkens,Ciobanu}. 
For either fragmented ($A_3>0$) or coherent states ($A_3<0$), 
a necessary condition besides $R$ being positive,  is that the 
shift \eqref{shift} fulfills $|\mathfrak{S}| \ll \frac N2$;  
in particular, a maximally fragmented state has $\mathfrak S=0$. 
In the continuum limit, using the Gaussian shape of the $\psi(j)$
distribution, it may be shown that 
the degree of fragmentation is given by 
\bea 
\mathfrak{F} =
\begin{cases}
1-\frac{|\mathfrak{S}|}{N/2} & \! A_3 \ge 0, \\  
1-e^{-\frac1{4N \sqrt R}}\sqrt{1-\left(
\!\frac{\mathfrak{S}}{N/2}\!\right)^2\!\left(1-e^{\,\frac1{2N\sqrt R}}\right)}
&\! A_3 <0. 
\end{cases}
\label{FragAnal} 
\ea
The above analytical approximation accurately reproduces the numerically 
obtained result, cf.\,\,Fig.\,\eqref{Fragment}.  
Note, 
provided we have the scaling
$\epsilon_1-\epsilon_0 = C  N^\gamma$ for large $N$, with 
$\gamma \le 1$ and $C$ constants,  assuming that all $A_i$ scale identically and
$A_1\neq A_2$ when $C\neq 0$, we conclude from  Eq.\,\eqref{shift} that  
fragmentation persists for $N\rightarrow \infty$. This is a 
remarkable feature of the present two-mode model. 

In case that $R$ becomes negative, or diverges, 
in the nondegenerate case $c_0 \neq c_N$ 
we simply obtain Fock states corresponding to the lower two-mode state, 
i.e. either $|N,0\rangle$ or $|0,N\rangle$. 
The degenerate case  
leads to coherent superpositions of macroscopically distinct quantum states,  
for which the original single peak in the $\psi_l$ distribution 
splits into two 
located around $l=0$ and $l=N$. 
The crossover is illustrated in Fig.\,\ref{cat}, 
by varying $A_4$ in \eqref{Rdef}, 
approaching the divergence of $R$. 
It appears that a fine-tuning of parameters is necessary, illustrating the
sensitivity of the superposed states to parameter fluctuations 
\cite{Cirac}.   
\vspace*{-0.5em}
\begin{center}
\begin{figure}[hbt]
\centering
\includegraphics[width=.36\textwidth]{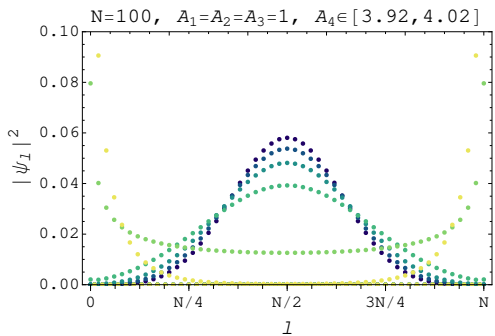}
\caption{\label{cat} Crossover to a coherent superposition of degenerate macroscopically distinct quantum states 
for diverging $R$ [Eq.\,\eqref{Rdef}], using a maximally fragmented state with ${\mathfrak S}=0$, 
upon variation of $A_4$ across the pole of $R$. Dark blue to light green 
(dark to light gray) indicates increasing $A_4$; 
$\Omega=0$.} 
\end{figure}\vspace*{-1.25em} 
\end{center} 
\vspace*{-0.5em} 

In conclusion, we have shown that 
adjusting the four interaction couplings in a two-mode model 
by employing the inhomogenity of the single-particle orbitals in a trap, 
in conjunction with a generally anisotropic and nonlocal interaction,  
leads to fragmented ground states whose degree of fragmentation is 
not sensitive to perturbations on the single-particle level. 
The physical origin of this 
interaction-based fragmentation is 
the pair-exchange process of bosons in a trap, not occurring in the continuum.  
Pair exchanges cause oscillations in the distribution of the many-body wavefunction amplitudes, which as a consequence 
entail vanishing first-order phase coherence
for positive pair-exchange coupling.

This research work was supported by the DFG under grant No. FI 690/3-1.


\begin{thebibliography}{9999}
\bibitem{Schroedinger} Such superposed macroscopic quantum 
states are sometimes referred to as ``Schr\"odinger cats'', e.g. in \cite{Ciobanu,Cirac}.
For clarity, we retain the explicit description of these states.
\bibitem{dipole-dipole} 
For an illustrative example, we use as the two states the ground and an 
excited state of a harmonic oscillator along $z$,
$\Psi_0(\rho,z)\equiv (a\sqrt{b}\,\pi^{\frac34})^{-1}
\exp\left[-{\rho^2}/{2 a^2} - {z^2}/{2 b^2}\right]$ and
$\Psi_1(\rho,z)\equiv (ab\sqrt{b/2}\,\pi^{\frac34})^{-1} 
z\exp\left[-{\rho^2}/{2 a^2} - {z^2}/{2 b^2}\right]$ 
(assuming cylindrical symmetry, where $a$ and $b$ are the 
half-axes of an ellipsoidal cloud \cite{Elgaroy}).
Writing the dipole-dipole interaction as $V({\bm r})
=\frac{3g_d}{4\pi |{\bm r}|^3} (1-3z^2/|{\bm r}|^2)$, where 
$g_d$ is the dipole coupling constant \cite{Fischer}, 
and adding the usual contact term, 
we obtain $A_1 = \frac{\sigma}{(\sqrt{2\pi} a)^3}
[g+g_d f_1(\sigma)]$, $A_2 =\frac34 \frac{\sigma}{(\sqrt{2\pi} a)^3}
[g+g_d f_2(\sigma)]$, $A_3 =\frac12 \frac{\sigma}{(\sqrt{2\pi} a)^3}
[g+g_d f_3(\sigma)]$, and $A_4 =2 A_1$,
where the $f_i(\sigma)$ are functions depending on the
aspect ratio $\sigma =  a/b$. 
They allow to sample the anisotropy 
and long-range character of the dipole-dipole 
interaction over the whole range from oblate ($\sigma >1$) to prolate 
($\sigma <1$), while with contact interactions only,  
the ratios of the $A_i$ are all constant \cite{Lahaye}. 
Explicitly, \sloppy
$f_1 = \frac{1}{\sigma^2-1} \left[2\sigma^2+1-3\sigma^2
\mathrm{ath}(\sigma) \right]$, 
$f_2 =\frac{1}{(\sigma^2-1)^3} \!
\left[2\sigma^6+4\sigma^4+8\sigma^2+1-\sigma^2(7+4\sigma^2
+4\sigma^4)\mathrm {ath}(\sigma)\right]$  and 
$f_3 =\frac{1}{(\sigma^2-1)^2}\left[2\sigma^4-10\sigma^2-1+9\sigma^2
\mathrm{ath}(\sigma)\right]$, 
where $\mathrm{ath}(\sigma) \equiv \mathrm{arctan}
[\sqrt{\sigma^2-1}]/{\sqrt{\sigma^2-1}}$.
\bibitem{note} We observe here that the authors of \cite{Elgaroy} were in fact 
considering what we call shifted coherent states, 
which have only one eigenvalue of the single-particle density matrix, 
instead of fragmented states.
\end{thebibliography}
\end{document}